%% file: hlis-digital-twin.tex
\documentclass[preprint,12pt]{elsarticle}

\usepackage{amssymb}
\usepackage{color}
\usepackage{amsthm}

\usepackage{multirow}

\usepackage{graphicx,subcaption}
\usepackage{amsmath}
\usepackage{amssymb}

\usepackage{lineno}

\usepackage{xcolor}
\usepackage{colortbl}
\usepackage{array}
\usepackage{pifont}
\usepackage{multirow}
\usepackage{hyperref}
\usepackage{framed}
\usepackage{changes}

\usepackage{tabularx}
\usepackage{enumitem}
\setlist{nolistsep}
\definecolor{green}{HTML}{66FF66}
\definecolor{novabarva}{HTML}{009700}

\journal{}

\begin{document}

\begin{frontmatter}

\title{Digital twins in sport: Concepts, Taxonomies, Challenges and Practical Potentials}

\author[inst1]{Tilen Hliš, Iztok Fister, Iztok Fister Jr.}

\affiliation[inst1]{organization={Faculty of Electrical Engineering and Computer Science, University of Maribor},
            addressline={Koroska cesta 46}, 
            city={Maribor},
            postcode={2000}, 
            country={Slovenia}}

\begin{abstract}
Digital twins belong to ten of the strategic technology trends according to the Gartner list from 2019, and have encountered a big expansion, especially with the introduction of Industry 4.0. Sport, on the other hand, has become a constant companion of the modern human suffering a lack of a healthy way of life. The application of digital twins in sport has brought dramatic changes not only in the domain of sport training, but also in managing athletes during competitions, searching for strategical solutions before and tactical solutions during the games by coaches. In this paper, the domain of digital twins in sport is reviewed based on papers which have emerged in this area. At first, the concept of a digital twin is discussed in general. Then, taxonomies of digital twins are appointed. According to these taxonomies, the collection of relevant papers is analyzed, and some real examples of digital twins are exposed. The review finishes with a discussion about how the digital twins affect changes in the modern sport disciplines, and what challenges and opportunities await the digital twins in the future.
\end{abstract}


\end{frontmatter}

\section{Introduction}
Nowadays, breaking the premise between the physical and virtual worlds can be done using digital twins. The Digital Twin (DT) concept is a replicated digital model of any device, product, or process. The history of DTs dates back in history, while, nowadays, they are used in the real world, where many applications were already developed. Indeed, DTs are, nowadays, applied to several areas, for example, the medicine domain~\cite{laubenbacher2024digital}, aerospace engineering~\cite{ferrari2024digital}, city planning~\cite{batty2024digital}, sport~\cite{lukavc2022digital}.

Since DTs are virtual replicas of physical assets, users can monitor real-time environments and perform various analyses. Data-making decisions are much more efficient, since they involve genuine, real-time data, which is crucial for efficient knowledge extraction from data. In many industries, it also allows users to design new products and digital replicas before producing a physical object, which also optimizes resources and costs. In Industry 4.0, and, later, in Industry 5.0, DTs can help optimize operations in production~\cite{wagner2019challenges,leng2021digital,lv2023digital}. Also, in the recent era of global warming, they can help reduce carbon footprints during the optimization of processes~\cite{ghita2021digital}.

Sports are physical activities that numerous people perform several days per week, primarily to maintain health, well-being, and relaxation. At the same time, countless serious athletes compete in different sports competitions~\cite{rauter2014mass}. Modern Information Technologies (IT) are also entering the world of sports and revolutionizing how people, athletes, and trainers perform different sports. From the simple wristwatches, which were capable of measuring the heart rate in the late 90s, then, later, with the smartwatches that could monitor more parameters~\cite{chandel2022smart}, e.g., GPS position, calories, and weather parameters, modern solutions emerged that can also give athletes advice during their training~\cite{kamivsalic2018sensors}. The main stepping stone lies in the data, which forms the way to make decisions~\cite{miller2013data}. Many decisions based on data are based on intelligent data analysis methods and tools, among which we can count Machine Learning (ML)~\cite{russell2016artificial}, Computational Intelligence (CI)~\cite{engelbrecht2007computational}, and various processes and techniques under the umbrella of data science~\cite{o2014data}. These methods can plan the sports training sessions~\cite{fister2019population}, analyze the sensor data arising during sports training/exercises~\cite{novatchkov2013artificial}, generate sports training routes~\cite{rajvsp2022modified}, etc. Not long ago, the concept of DT was also introduced into sports in different realms, e.g., as the role of a personal assistant during sports training sessions~\cite{lukavc2022digital}.

In this paper, we take a closer look at the DTs arising in a vibrant area of sports, sports training, human movement, and competition. Our goals are to review the current state of this research area, propose a taxonomy of DTs in sports, identify the main challenges and obstacles of this technology, and produce new guidelines for creating new DTs.

This review article also builds on knowledge from previous review articles in this field, where the authors have captured DTs in sports and attempted to construct new knowledge mosaics. The following review papers~\cite{gamez2020digital,pascual2023systematic} are the closest to our review paper. Indeed, our paper tackles the actual state of DTs in sports, and proposes a new taxonomy that should be a stepping stone for developing new DTs in sports.

The contributions of this paper are as follows:
\begin{itemize}
    \item an overview is conducted of DTs in sport,
    \item a taxonomy is proposed for DTs in sport,
    \item the main DTs are identified in theory and practical use,
    \item future challenges are studied and opportunities for further use.
\end{itemize}

The structure of the remainder of paper is as follows: Section~\ref{sec:2} reveals the fundamentals of DTs. The research methodology for assembling a collection of papers for extensive analysis is discussed in Section~\ref{sec:3}. In Section~\ref{sec:4}, a detailed analysis of DTs in sport is described, while the taxonomies of DTs in sport are appointed in Section~\ref{sec:5}. Real examples of DTs in sport are the subject of Section~\ref{sec:6}. Section~\ref{sec:7} deals with the question of how the DTs change the modern sport disciplines. Challenges and opportunities are treated in Section~\ref{sec:8}, while the paper is finished with Section~\ref{sec:9}, where also the directions are outlined for future work. 

\section{Fundamentals of digital twins}\label{sec:2}
A DT is a replicated digital model of any physical entity (i.e., device, product, or process) with the same behavior as the original ones. In Gartner's list from 2019, it belongs to the top ten strategic technology trends. The DT roots originated in 1960 from the National Aeronautics and Space Administration (NASA) Apollo program~\cite{Romero2020}, within which space flights were simulated in the virtual environment during the various astronauts' training phases. With the advent of Industry~4.0 and Industry~5.0, the technology was integrated into unified models (i.e., smart factories) that drive product design, manufacturing, and cyber-security~\cite{chen2021digital,holzinger2024human}.

\begin{figure}[htb]
    \centering
    \includegraphics[scale=0.6]{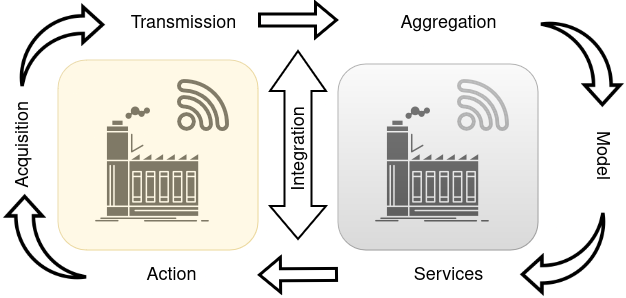}
    \caption{Concept of a Digital Twin.}
    \label{fig:DT}
\end{figure}

A concept of a DT is illustrated in Fig.~\ref{fig:DT}, from which it can be seen that the DT integrates the physical and digital world with a complex data manipulation. Data which arise during the data acquisition phase by a physical entity are transmitted to its virtual counterpart. 
Before being used by the DT processing model, they need to be fused with data from external sources (e.g., external databases, or the Internet), and cleaned during an aggregation phase. Indeed, the DT processing model models the properties of the physical object. The majority of the DT applications needs a lot of services (e.g., Association Rule Mining (ARM)) that are incorporated into a DT processing model. The results of the DT processing model are represented in the action phase in different forms that serve to help decision-makers by accepting the complex decisions.

\begin{figure}[htb]
    \centering
    \includegraphics[scale=0.75]{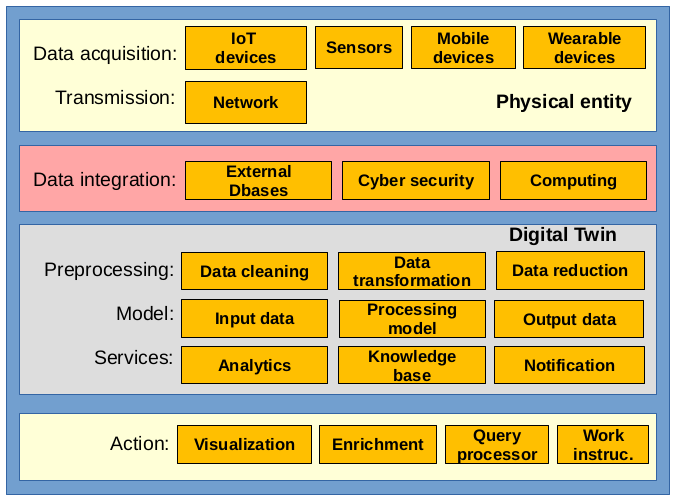}
    \caption{Architecture of a Digital Twin.}
    \label{fig:DT-2}
\end{figure}

The architecture of a DT consists of more components, working interconnectedly between each other, which, thus, determines the workflow of data across those. The components of the typical DT are as follows (Fig.~\ref{fig:DT-2})~\cite{Patel2022insight}:
\begin{enumerate}
    \item data acquisition,
    \item transmission,
    \item data integration,
    \item data preprocessing,
    \item processing model,
    \item services,
    \item actions.
\end{enumerate}
Interestingly, the first two components are devoted to sensing the physical world, while the last one affects it with actions. The third component integrates the physical world with the virtual one, while the next three represent an implementation of the specific DT. 

The data acquisition component allows acquiring data from various Internet of Things (IoT) devices, sensors, mobile devices, and wearable devices, with which a physical entity is monitored. The acquired data are then transmitted to the DT using network services. 

In general, the data integration component addresses problems of fusion of the acquired data from the physical world with data from external databases via different network interfaces. This component is devoted also for protecting the computing devices and sensitive data from ransomware attacks~\cite{joshi2022digital}. 

The data preprocessing component provides that unstructured data from IoT sensors, as well as other structured and unstructured data from other external sources are transformed into a format, which can be understood and analyzed by computers and ML methods. In line with this, more data preprocessing methods can be applied, e.g., data cleaning, data transformation, and data reduction. The model consists of three sub-components, as follow~\cite{joshi2022digital}:  (1) input data, (2) DT process model, and (3) output data. The input data is responsible for preparing unstructured data, either acquired from various sources, like IoT devices and sensors, or structured data from external databases in a format understandable by the DT process model. The DT process model is a representation of a physical entity, and, therefore, the model must be able to simulate its operations. This model produces output data similar to the physical entity. There are especially three kinds of services used by a DT process model: (1) analytics, (2) knowledge base, and (3) notification. Analytics refers to predetermined ML models that are often used for forecasting or anomaly detection~\cite{joshi2022digital}. Knowledge base addresses those forecasting ML models that reports their output through a set of rules. Notifications take place when the outcome of analytics needs attention and can be communicated using different mechanisms (e.g., e-mail, SMS, etc.). 

A DT user is notified about performing a potential action through application interfaces. Typically, the following application interfaces are applied more frequently a by DT: (1) visualization, (2) enrichment, (3) query processor, and (4) work instructions. Visualization is a tool often used for gaining insights into the current status of the particular DT, and thus enables a user to respond to the situation properly. An enrichment changes the way in which the results of the DT model made in the virtual environment interact with the physical world. The query processor application interface allows the DT user to enter a query to the specific data source, and the obtained results are applied in the decision-making process. Work instructions are an application interface capable of instructing the DT user by a description of specific actions, typically, in the sense of the AR. 

\subsection{Digital twins in sport}
Sport plays an important role in human life. Moreover, modern sport can be treated as a surrogate religion or popular theater, in which people identify themselves with the sport's champions~\cite{fister2019computational}. On the other hand, the modern sedentary lifestyle, and, consequently, the lack of activity cause obesity and loss of fitness by modern people. The lack of activity hurt them the most in the time of the Corona19 pandemic. The crisis had an especially bad effect on the younger generation. In line with this, the sport can help to improve those harmful influences on the modern lifestyle.

Recently, DT technology has become increasingly significant in the sport domain, where it provides advanced solutions for sport training, performance monitoring, and strategical planning~\cite{miller2022unified}. Indeed, the DT in sport presents a kind of human DT, where the human (i.e., the athlete) is placed in the center of the scene. Thus, humans are monitored in the physical world by various wearable devices and sensors, while the data acquired from these devices are transmitted to both, i.e., the real athletes and their virtual replicas modeled by the DT. These data are analyzed using the DT process model, while the obtained results are passed to the monitored athlete in the sense of an Augmented Reality (AR). The athlete then makes a decision on how to respond to a specific situation during a training session or competition. Interestingly, the DT in sport does not distinguish between the professional and amateur athletes.

\section{Research Methodology}\label{sec:3}

The goal of this review was to: (1) explore how digital technology is being integrated into sports training to enhance the capabilities of athletes and coaches, and (2) assess the extent to which these technological innovations are being implemented in real-world sports settings. Based on these objectives, the following Research Questions (RQ) were formulated:

\begin{itemize}
    \item \textbf{RQ1}: Which sports are supported the most by DTs?\label{rq1}
    \item \textbf{RQ2}: What is the level of maturity and practical implementation of DT technology in sports settings?\label{rq2}
    \item \textbf{RQ3}: How do DTs influence sports training?\label{rq3}
    \item \textbf{RQ4}: What are the challenges of implementing DTs in sports?\label{rq4}
\end{itemize}

A \textbf{Systematic Literature Review (SLR)} was the empirical research method to answer the research questions. The following SLR Guidelines in Software Engineering~\cite{Kitchenham2007} were followed to conduct this review by us: We initially scanned the domain by examining the relevant literature from leading digital databases in software engineering, prepared the SLR (conducted from April 15 to April 23, 2024) and developed appropriate search strings. During the initial domain scan, it became clear that actual implementations of DTs in sports are notably rare. There are only a few studies focused specifically on this area within sports. In contrast, most research on human DTs focuses predominantly on healthcare. Additionally, while DTs originate from industrial applications and are utilized extensively in sectors such as logistics and manufacturing \cite{Jiang2021}, their adaptation to human-centric applications outside of healthcare, especially in sports, remains limited.

We formed a search string based on two groups of keywords. Group \textbf{one} included the keywords "digital twin" (variation: "digital twins"). Group \textbf{two} included "sport" (variation: "sports"), "fitness", "coaching", and "virtual trainer". Based on the keywords, the single aggregated search string used to perform the SLR was as follows:
\begin{center}
    \footnotesize
    ("digital twin") \textbf{AND} ("sport" \textbf{OR} "fitness" \textbf{OR} "coaching" \textbf{OR} "virtual trainer")
\end{center}

Variations in the search strings used across different databases were necessary, due to the distinct query languages and constraints unique to each scientific paper database. The queried databases are shown in Table \ref{tab:search_results}.

\begin{table}[ht]
\centering
\caption{Databases with search results and total after removing duplicates.}
\begin{tabular}{llrr}
\hline
\multirow{2}{*}{\textbf{Database Name}} & \multirow{2}{*}{\textbf{URL}} & \multicolumn{2}{c}{\textbf{Number of papers}}\\\cline{3-4}
 & & \textbf{Total} & \textbf{Included} \\
\hline
ACM Digital Library & dl.acm.org & 38 & 0 \\
Google Scholar & scholar.google.com & 414 & 15 \\
IEEE Xplore & ieeexplore.ieee.org & 92 & 1 \\
ScienceDirect & sciencedirect.com & 9 & 0 \\
Scopus & scopus.com & 86 & 5 \\
SpringerLink & link.springer.com & 100 & 3 \\
\hline
\multicolumn{2}{l}{\textbf{Total}} & 739 & 24 \\
\hline
\end{tabular}
\label{tab:search_results}
\end{table}

The selection and exclusion criteria were defined clearly, and the limitations were considered carefully, to ensure a comprehensive and current understanding of the field. 

Thus, the \textbf{selection criteria} were as follows:
\begin{itemize}
    \item The research focused specifically on implementing or applying DT technology in sports for athletes or coaches.
    \item The research underwent a peer-review process.
    \item The study pertained to sports as athletic activities demanding skill, physical capability, or competitiveness.
    \item The research involved using DT technology or related computational methods, such as simulation modeling, artificial intelligence, or real-time data analysis, in the context of sports.
\end{itemize}

We considered the following \textbf{exclusion criteria}:
\begin{itemize}
    \item The research was not in English.
    \item The full text of the research was not accessible through the digital library or any subscription services.
    \item The research focused solely on recognizing activities from a leisure perspective, such as general health.
\end{itemize}

\textbf{The limitations} of the paper selection were as follows: 
\begin{itemize}
    \item The research was confined to six scientific databases/search engines: ACM Digital Library, Google Scholar, IEEE Xplore, ScienceDirect, Scopus, and SpringerLink.
    \item The research needed to be published before April 15, 2024, which is when the indexing of potential articles took place.
\end{itemize}

The \textbf{SLR Progression} is shown in Fig.~\ref{fig:progression}:

\begin{itemize}
    \item \textbf{Phase 1: Initial search.}
    In the first phase, we conducted an initial search, which produced \textbf{739} results. Out of these, only \textbf{493} were available to be read in full-text form. During our preliminary research, we discovered that most works concentrate solely on the health aspect rather than sports, which led us to add an exclusion criterion: research focused exclusively on recognizing activities from a leisure perspective, such as general health.

\begin{figure}[htb]
    \centering
    \includegraphics[scale=0.22]{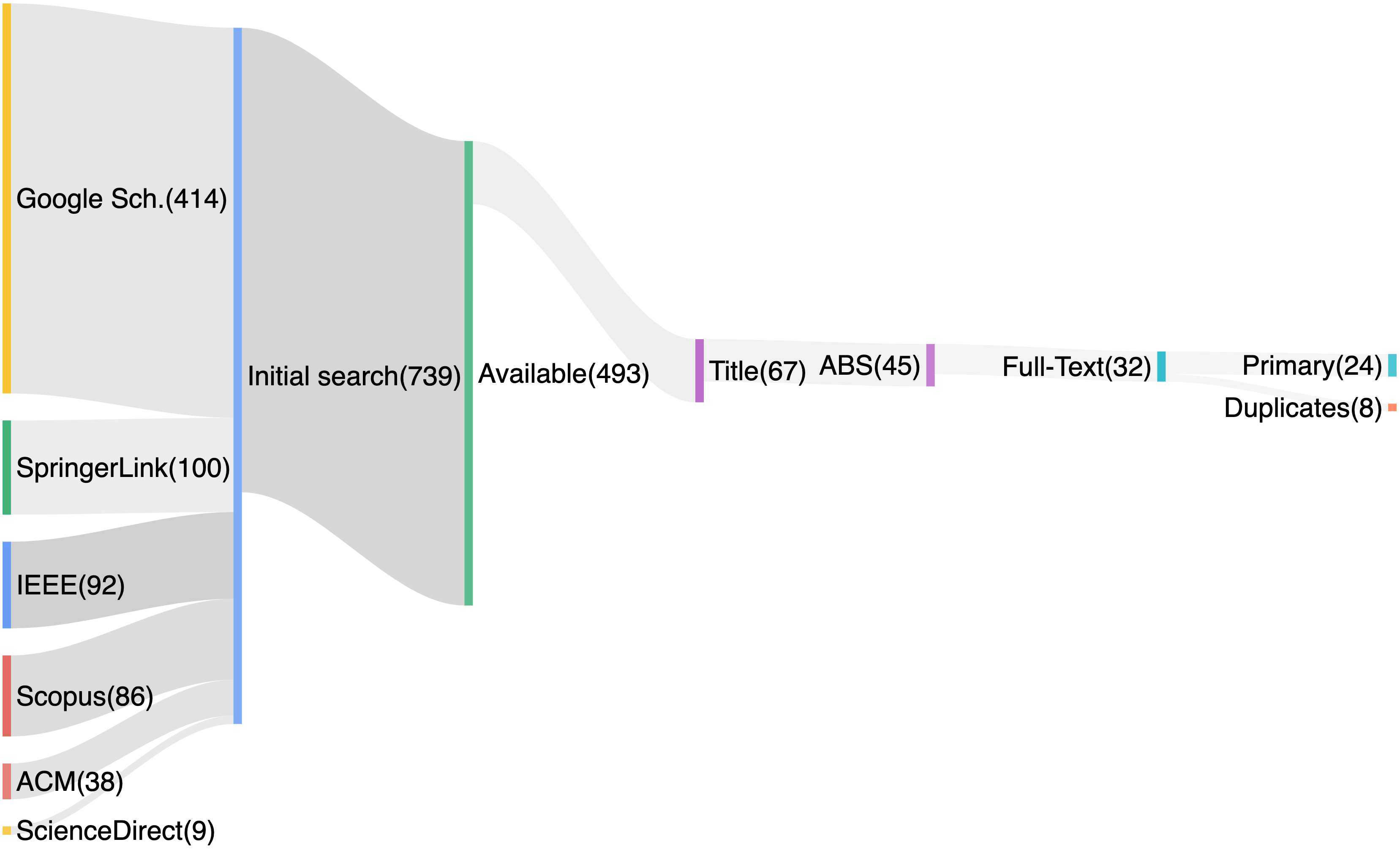}
    \caption{SLR Progression}
    \label{fig:progression}
\end{figure}
    
    \item \textbf{Phase 2: Title-based screening.}
    In this phase, three researchers reviewed the study titles independently. Their findings were combined to promote a well-rounded and impartial selection process. This step allowed us to pinpoint studies that were appropriate for the next stage of the review (\textbf{67} results).
    
    \item \textbf{Phase 3: Abstract-based screening.}
    In this phase, the abstracts and keywords of each study were reviewed carefully. The results were then aggregated, to identify studies that met the criteria (\textbf{45} results), ensuring a focused selection process for the next review stage.    
    
    \item \textbf{Phase 4: Full-text review.}
    The full-text review was the subsequent phase of the SLR, yielding an initial set of 32 primary studies. After removing duplicates, we had a final collection of \textbf{24} primary studies.
    
    \item \textbf{Phase 5: Snow-balling.}
    In the final phase, we used a related work review to identify additional studies through the snowballing technique. However, no new literature was found during this process, so the review remained focused on the \textbf{24} primary studies identified in the previous phase.
\end{itemize}

The results of the literature search analysis are provided in the remainder of the paper.

\section{Detailed analysis of the DT in sport}\label{sec:4}
The purpose of the analysis was to evidence details of the various DTs that were found in the collection of the papers, and to expose their main characteristics. The results of the investigation are summarized in Table~\ref{tab:detail} that, in columns, presents information like: (1) the name of the specific DT, (2) the data acquisition methods for exploring the physical world, (3) the DT model for simulating the real entity, and (4) the references to the papers where these were described, according to the particular sports discipline.

\begin{table}[htb]
    \caption{Detailed analysis of the DT in sport.}
    \label{tab:detail}
    \centering
    \scriptsize
    \begin{tabular}{llllr}
         \hline
         Sport & Name & Data acquisition & DT model & Papers \\
         \hline\hline
          & DT Coaching & Video cameras & OpenPose & \cite{chen2022digital} \\ 
         Fitness \& & DTCoach & Cameras & Shallow NN & \cite{díaz2021dtcoach,gamez2020digital,gamez2021digital} \\ 
         Health & Digital Athlete & Depth Cameras & NMSK, LSTM & \cite{lloyd2023maintaining} \\
          & SmartFit & IoT devices & ECOC & \cite{barricelli2020human} \\
          & DT for Fitness & IoT devices & MLSTM & \cite{alsubai2023hybrid} \\
          \hline
         \multirow{3}{*}{Cycling} & AST-Monitor & ANT+ sensors & Real-time analysis   & \cite{lukavc2022digital,fister2021trainer,fister2021ast} \\
          & \multirow{2}{*}{DT Model} & Cycling ergometer, & \multirow{2}{*}{Margaria-Morton model} & \multirow{2}{*}{\cite{boillet2024individualized}} \\
          &  & instrumented pedals &  &  \\ \hline
          \multirow{4}{*}{Football} & \multirow{2}{*}{Conn. Footballer} & Chest strap, & \multirow{2}{*}{MC Simulation Algorithm} & \multirow{2}{*}{\cite{balachandar2019}} \\
           & & RFID tag & &  \\
          & Athlete Training Sys. & Smart insoles & Probabilistic Model & \cite{laamarti2019towards} \\          
          & Aid Robots DT & Lidar, Camera & Dec-POMDP & \cite{pham2023decision} \\   & Turtle DT & Cameras & Real-Time Database & \cite{walravens2022soccer} \\ 
         \hline
       \multirow{3}{*}{Shooting} & \multirow{3}{*}{Shooting DT} & Wearable sensors, & \multirow{3}{*}{Decision Tree Algorithm} & \multirow{3}{*}{\cite{morzenti2023human}} \\
        & & motion capture,  & &  \\
        & & physiological sensors & &  \\
          \hline
         \multirow{1}{*}{Education} & DT in education & Kinect sensors & Hidden Markov Model & \cite{liu2022digital} \\
          \hline
         \multirow{1}{*}{Dancing} & MetaHuman & Dance motion capture & OpenSim & \cite{cedermalm2022simulating} \\
          \hline
         \multirow{1}{*}{Swimming} & DT for swimmers & IMU, Force sensor & Newton’s laws of motion & \cite{douglass2024swimming} \\
          \hline
         \multirow{2}{*}{Gymnastics} & \multirow{2}{*}{DT for Gymnastics} & VR image recognition,  & Motion sequence generation  & \multirow{2}{*}{\cite{shi2021application}} \\
          & & motion sensors & motion editing algorithm &  \\ \hline\hline 
    \end{tabular}
\end{table}

The following conclusions can be exposed from Table~\ref{tab:detail}: The crucial differences in DT implementations have been indicated by application in different sports. Thus, the biggest differences arise when comparing the implementations of DTs supporting either individual or team sports. In individual sports, the DTs play the role of training assistant, virtual coach, or even educator. As a training assistant, the DTs try to enhance an athlete's performance and reduce their injury risks. The virtual coach DT simulates the tasks of real coaches, and, thus, tries to replace their role in the training process. Educator DTs replace either the real educators in physical education or teachers of artistic gymnastics. Usually, these application solutions are equipped with IoT devices, RFID tags, and wearable non-invasive sensors (e.g., ANT+ sensors). Recently, the DTs have been equipped with video cameras (e.g., DTCoach) and Kinect or motion sensors (e.g., in gymnastics, dancing, and education), where they try to improve the athlete's movement.

On the other hand, the DTs, in team sports like football and basketball, serve as real coaches for improving strategic and tactical decisions or monitoring and simulating players' movements and health conditions in real time. Typically, these solutions explore the power of smart insoles and haptic feedback devices (e.g., a smart haptic display).

Different NNs are applied for modelling the physical world. Thus, the shallow NN with only one or two hidden layers and an LSTM for dealing with data based on time series were discovered during the analysis. The primary characteristic of these NNs is the time complexity, and, consequently, these are not the most appropriate for use in real-time applications. Therefore, the more appropriate models are based on real-time analysis, MC simulation, and Hidden MM. Some models were developed especially for particular sports disciplines, for instance, motion sequence generation in gymnastics or Newton's laws of motion in swimming. 

\section{Taxonomy of the digital twin in sport}\label{sec:5}
The characteristics of various kinds of DTs were, theoretically, well elaborated~\cite{Patel2022insight}. Therefore, the taxonomy of DT in sport follows these hints. Indeed, the classification of DTs can be performed by considering more aspects, such as, for instance:
\begin{itemize}
    \item types of DTs,
    \item applications of DTs,
    \item traits of DTs.
\end{itemize}
In the remainder of the paper, the exposed aspects are discussed in more detail.

\subsection{Types of Digital Twins}
Grieves, one of the pioneers of DT development, in his early study from 2015~\cite{grieves2014digital}, established that a DT consists of three components, as follows: (1) a physical product, (2) a virtual representation of the product, and (3) the bi-directional data connection from replicas of the physical entity to a virtual representation, as well as information and processes from the virtual representation, to a replica of the physical entity. These components form the so-called product life-cycle that can also be applied to DT development. Later, Grieves in~\cite{Grieves2017digital} defined three types of DTs entering into the product life-cycle, as follows:
\begin{itemize}
\item Digital Twin Prototype (DTP): A \textit{prototype} that served as the initial model for the DT, often used in the design and development phase (i.e., system shape of the DTP in the virtual world only).
\item Digital Twin Instance (DTI): An \textit{instance} representing a specific, individual unit of equipment or system in operation that connects the virtual and physical worlds.
\item Digital Twin Environment (DTE): An \textit{aggregate of multiple DTIs or DTPs} used for analyzing trends and performance across a fleet or system. The DTEs are categorized based on their functionality, as:
\begin{itemize}
\item Predictive: These DTs forecast future behaviors and performances using historical and real-time data, aiding in preemptive decision-making and strategy development.
\item Interrogative: These systems focus on displaying the current and past states of the physical system, enabling users to query and understand historical performance and operational metrics.
\end{itemize}
\end{itemize}
Actually, a DT model goes through three phases during its life-cycle~\cite{Grieves2017digital}: At the beginning, it emerges virtually as a prototype, then, continues through its operational life, and, finally, it is eventually retired and disposed of.

\subsection{Applications of Digital Twins}
The area of DT applications depends on the level of product magnification. Thus, different types of DT applications can co-exist within a system or process. In general, regarding the type of applications, DTs can be classified as follows~\cite{Patel2022insight}:
\begin{itemize}
\item Component: A digital model of individual components or parts within a larger system.
\item Product: A digital model of entire products or assets capable of simulating their behavior in different conditions.
\item System: A digital model of entire systems, offering insights into complex interactions.
\item Process: A digital model of entire business processes used often used in manufacturing or operational contexts.
\end{itemize}
As can be seen from the classification, the area of application can lead to big differences in the implementation of the system.

\subsection{Traits of Digital Twins}
Traits of DTs refer to how similar/different are the characteristics of the implemented virtual replica when comparing it with the physical entity. 
The following traits of DT are found, as follows:
\begin{itemize}
\item Same look as the original entity: The DT's look is the same as that of the original entity.
\item Different details of the original entity: The DT has the look augmented with different details of the original object, with which they supplement the total look of the effective replica.
\item The same behavior as the original entity: This trait enables a DT to behave and look the same as the original entity.
\item Prediction and information about problems that could arise in ad-vance: DTs with these traits are able to predict the problems that could occur in the future, and, consequently, try to avoid them.
\end{itemize}
In general, human DTs in sport do not have the same look as the original ones, but operate with different details of the original. In other words, these DTs operate with the physical information that determines the particular human in the real-world, and, thus, mimic their behavior.

\subsection{Summary}
The proposed taxonomy of the DTs in sport classifies the collection of elaborated papers according to three aspects, as follows: (1) types, (2) applications, and (3) traits of the DT. The results of the investigation are illustrated in Table~\ref{fig:taxonomy}, that is organized as a 3D cube, where the different DT aspects are represented on the X-axis, the names of the DTs on the Y-axis, and attributes of the aspects on the Z-axis. Interestingly, the number of papers, in which the definite DT is mentioned, has been added to the "Type" aspect.
\input{featuretable}

As can be seen in Table~\ref{fig:taxonomy}, the majority of the DTs are described by only one paper. This could be evidence of a fact that these DTs are either in the prototype phase, or have already come out of development phase, and, therefore, need additional experimental work to get proper scalability and efficiency for using in practice. The exceptions are AST Monitor and DTCoach that are introduced in three papers, which represent the development of both from the DTP to DTE. 

The area of applications for the different DTs in the Table started from the component via product to system. No one of the observed DTs is a part of the big process. The AST Monitor is a part of the universal AST~\cite{fister2021trainer}, while the DTCoach bases on the pre-trained light pose estimation model capable of feedback generation using a mobile device, and can be applied for many fitness, health and well-being purposes. On the other hand, the Athlete Training System consists of more different DTEs that can be applied in the same application areas supported by DTCoach. The majority of the remaining proposed DTs emerged as components that are devoted to train only a particular training area in a specific sport discipline. For instance, the shooting DT~\cite{morzenti2023human} focuses on monitoring the shooter's pose by measuring kinematics and physiological sensors. Thus, the shooter is only one of the components during this training session. Well-represented are also DTs covering the product application area. Digital Athlete~\cite{lloyd2023maintaining}, for example, concentrates itself on maintaining a soldier's muscolo-skeletal health with wearable sensors and computer vision, and, thus, represents the complete solution of this military area.

The majority of the analyzed DTs have a look with different details of the original human, but all except DTPs are able to make predictions about problems that could arise in advance. 

\section{Real examples of digital twins}\label{sec:6}

DTs are revolutionizing sports by enhancing training and performance analysis with advanced data-driven insights. This technology now provides athletes and coaches with precise simulations and real-time feedback, fundamentally transforming athletic preparation and strategy. 

The aim of the study was to analyze two of the more complete real-world solutions within the DT system, as follows:
\begin{itemize}
    \item AST-Monitor,
    \item DTCoach.
\end{itemize}
Both real-world implementations cover an area of sport training, but in different sport disciplines, i.e., cycling and fitness. Their completeness is also evident by more papers being published that highlight these DT systems from various aspects. 

In the remainder of the section, the mentioned real-world implementations are explored across various sports disciplines in more detail. The section is finished with an analysis of the key factors for estimating the quality of DT implementation.

\subsection{AST-Monitor}

\begin{figure}[htb]
    \centering
    \includegraphics[scale=0.75]{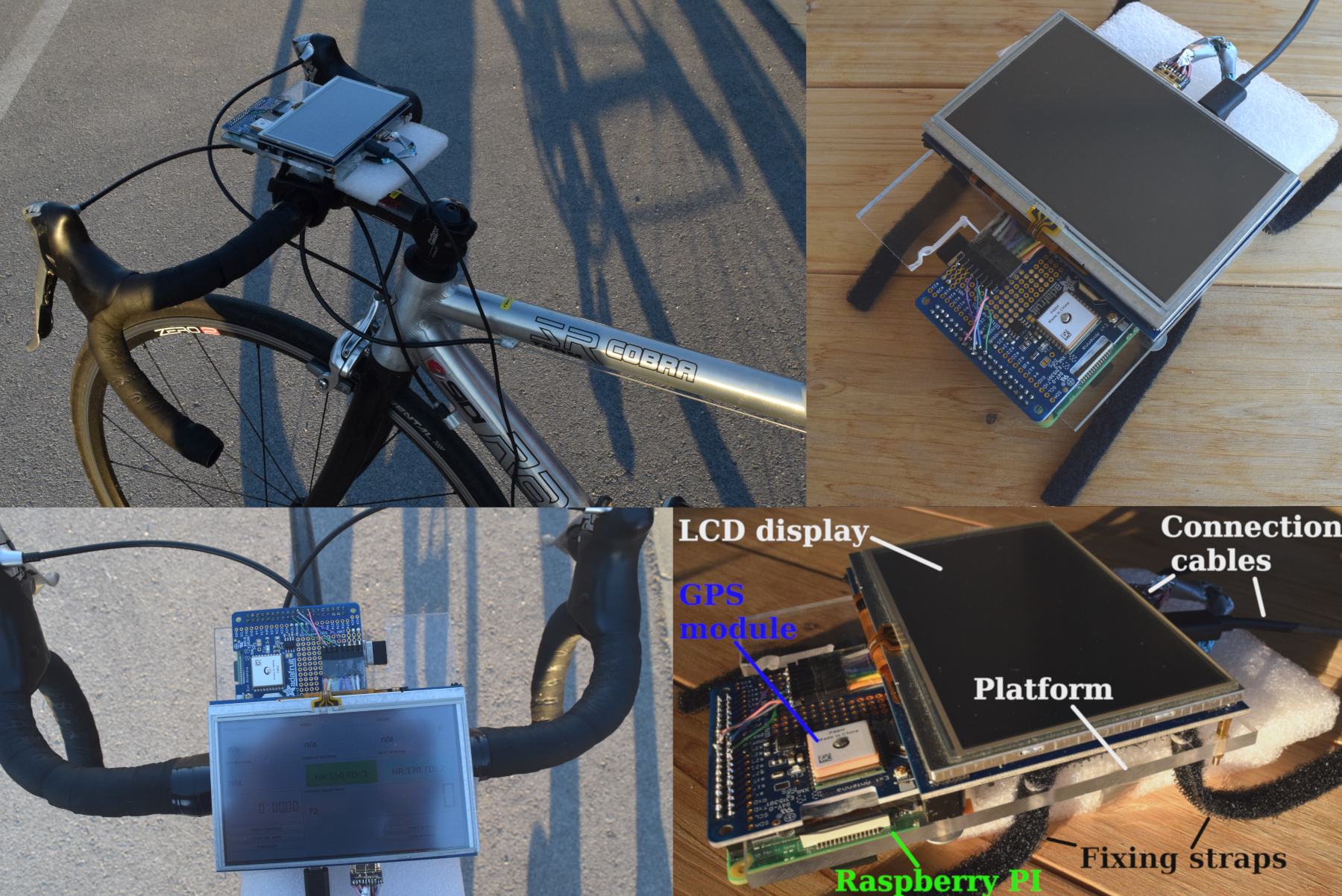}
    \caption[AST-Monitor]{AST-Monitor \footnotemark }
    \label{fig:AST-monitor}
\end{figure}

AST monitor (Fig.~\ref{fig:AST-monitor}) is a part of the universal Artificial Sport Trainer (AST)~\cite{fister2021ast} dedicated for monitoring an athlete during the cycling training session. A specification of the training session is obtained from the AST system in the form of a Domain-Specific Language (DSL). The DT advises the cyclist during the training session with personalized guidance in the form of an AR via display connected to Raspberry Pi microcomputer, which is mounted on the cycle. A powerbank is used as a source of electrical energy. A DT model is implemented as a feedback system based on the simple mathematical model for prediction of the Heart Rate (HR), necessary to satisfy the requirements of the training session by the cyclist. Among other features, the AST Monitor is equipped with a variety of ANT+ sensors connected to the Raspberry Pi, like HR, power meter, and speed sensor, as well as a GPS receiver and a WiFi transmitter. The Raspberry Pi enable mobility, scalability, security, and connectivity, and, therefore, represents a promising platform for development of DTs in the future.

\footnotetext{Fister, Jr., I., \& Fister, D. (2024). firefly-cpp/figures: 1.0 (1.0). Zenodo. https://doi.org/10.5281/zenodo.10479320}

\subsection{DTCoach}

In the study from Diaz \cite{díaz2021dtcoach}, a DTI is developed for real-time physical activity coaching on mobile devices, highlighting an interrogative DTE that provides immediate feedback based on user performance. This system utilizes edge computing to enable a portable and accessible coaching product that operates effectively on users’ smartphones. The DT features advanced machine learning algorithms, including lightweight neural networks for pose estimation, tailored to run efficiently on devices with limited processing power. This capability ensures that users receive personalized training guidance directly from their mobile devices without the need for complex sensors or additional hardware. The system's main advantages include its ability to offer real-time, personalized feedback, enhancing the effectiveness of home-based physical training significantly. Implementing this technology posed several challenges, particularly in optimizing data processing to function within the constraints of mobile technology. The solution employed pre-processing techniques to reduce computational demands, and ensure the system's responsiveness. Additionally, the development team faced challenges related to ensuring data security and user privacy, critical when handling sensitive health data on mobile platforms. This DT application not only supports individualized fitness goals, but also adapts to each user's unique physical conditions, providing a scalable solution that enhances physical activity through advanced technological integration \cite{díaz2021dtcoach}.

\subsection{Key factors for estimating the quality of Digital Twin implementations}
The key factors that define the capabilities and benefits of DTs were estimated according to the following aspects:
\begin{itemize}
\item Capability: What the DT is technically capable of doing.
\item Advantages: The specific benefits provided by the DT, such as increased efficiency or reduced downtime.
\item Features: Distinctive attributes or functions of the DT system.
\end{itemize}
The results of analyzing the key factors for estimating the quality of DT implementation of AST Monitor and DTCoach can be observed in Table~\ref{tab:estimation},
\begin{table}[htb]
    \caption{Estimation of the quality of real-world DT implementations.}
    \label{tab:estimation}
    \small
    \centering
    \begin{tabular}{lll}
        \hline
        Key factor & AST-Monitor & DTCoach \\
        \hline\hline
        \multirow{2}{*}{Capability} & Personalized training guidance & Personalized training guidance \\ 
         & from RaspBerry Pi &  from a mobile device \\
        Advantages & Real-time AR feedback & Real-time personalized feedback \\ 
        \multirow{3}{*}{Features} & Lightweight prediction algor., & Lightweight NN, limited \\
         & Variety ANT+ sensors,  &  processing power \\
         & Raspbery Pi & \\
        \hline
    \end{tabular}
\end{table}
from which it can be indicated that users of the former interact with the DT model using Raspberry Pi, while users of the latter use mobile devices.

\section{How digital twins shape the world of current and future sports?}\label{sec:7}

DTs offer new potential in sports by creating virtual replicas of athletes in their environments. Our extensive literature review discovered some areas that need to be exposed. For example, these DT replicas can be applied in the following sport domains: 
\begin{itemize}
    \item performance analysis and enhanced training,
    \item injury prevention and recovery,
    \item strategic planning and competition simulation.
\end{itemize}
The results of the investigation study are summarized in Table~\ref{tab:benefits}, that illustrates the key benefits of DTs in sport, which could change the current and future sport radically.
\begin{table}[htb]
    \caption{Key benefits of digital twins in sports.}
    \label{tab:benefits}
    \small
    \centering
    \begin{tabular}{lll}
        \hline
        Domain & Key benefits & Papers \\
        \hline\hline
        Performance analysis & Personalized training, & \cite{lukavc2022digital, gamez2020digital, barricelli2020human, fister2021trainer, balachandar2019} \\ 
        and enhanced training & real-time adaptations & \cite{laamarti2019towards, morzenti2023human, douglass2024swimming, sahal2022personal} \\
        \hline
        Injury prevention & Biomechanical analysis, & \cite{gamez2020digital, lloyd2023maintaining, laamarti2019towards, lauer2024towards} \\ 
        and recovery & real-time recovery monitoring & \\
        \hline
        Strategy planning and  & Data-driven insights, & \cite{pham2023decision, walravens2022soccer} \\
        competition simulation & real-time decision-making  &  \\
        \hline
    \end{tabular}
\end{table}
In the remainder of the paper, the key benefits of DTs in sports are dealt with in detail.

\subsection{Performance analysis and enhanced training}

DTs integrate data from various sensors (e.g., IoT devices, wearable sensors) to monitor biometrics such as heart rate, oxygen levels, and muscle activity~\cite{lukavc2022digital, gamez2020digital, laamarti2019towards}. These data allows highly personalized training programs that adapt to the athlete's needs and conditions in real-time~\cite{balachandar2019, morzenti2023human}. 

The development of smart training systems gives us advanced simulations that help correct technical errors and optimize movements. For instance, DTs in cycling can simulate interval training scenarios, to advise the best strategies for improving efficiency and performance~\cite{lukavc2022digital}.

Another example presents sports like swimming, where DTs analyze underwater movements, in order to optimize techniques, such as streamline and dolphin kick, leading to significant performance gains~\cite{douglass2024swimming}. 

One key aspect of training is psychological and motivational support, achieved by visualizing progress, setting realistic goals, and providing continuous feedback. This helps maintain an athlete's motivation and adherence to training programs. On the other hand, they can simulate the psychological impact of different training and competition scenarios, helping athletes prepare mentally for high-stress situations~\cite{morzenti2023human, sahal2022personal, barricelli2020human, fister2021trainer}.

\subsection{Injury prevention and recovery}

As many authors have stated, DTs enable detailed analysis of an athlete's biomechanics, help to identify potential injury risks, and correct his/her wrong techniques. For example, the use of sensors to capture detailed motion data and assess techniques, can pinpoint areas that need improvement and prevent injuries~\cite{gamez2020digital, lloyd2023maintaining, laamarti2019towards}. 

On the other hand, rehabilitation is an important aspect that needs to be covered. DTs assist rehabilitation by monitoring recovery progress and adjusting training loads accordingly. They can simulate recovery processes and design optimal rehabilitation protocols to ensure athletes return to their peak performance safely~\cite{laamarti2019towards, gamez2020digital, lauer2024towards, lloyd2023maintaining}.

\subsection{Strategy planning and competition simulation}

DTs enable athletes and coaches to simulate various competition scenarios and strategies, especially in team sports, providing a powerful tool for optimizing performance. By creating virtual replicas of different game situations, DTs offer data-driven insights that help understand the best performance-enhancing approaches. This detailed simulation allows coaches of teams to make informed decisions regarding the strategy and tactics of the ongoing games. In line with this, the coaches can find the proper position for the player in the game, and, thus, they can ensure that every player's move is planned strategically and executed based on thorough analysis~\cite{walravens2022soccer}. 

Moreover, DTs also play a crucial role in real-time decision-making in team sports like football. They simulate game scenarios as they unfold, allowing coaches to adjust strategically during matches. These real-time simulations are driven by predictive analytics, which analyze ongoing gameplay to forecast potential outcomes and suggest optimal strategies. This capability ensures that coaches can respond swiftly to changing conditions on the field, making decisions that can impact the game's outcome significantly. By leveraging the power of DTs, teams can enhance their tactical understanding, and stay ahead in competitive sports environments~\cite{pham2023decision}.

\section{Challenges and opportunities for future development}\label{sec:8}
Advantages in DT technology have connected seamlessly with growing challenges in areas like AI and IoT. Indeed, implementing DTs involves the following challenges that need to be addressed:
\begin{itemize}
\item technical and setup challenges, 
\item expertise requirements, 
\item data security and privacy. 
\end{itemize}
Technical and setup challenges include data acquisition, sensor setup, and creating a seamless link between the real and virtual worlds. Fault tolerance in sensors is crucial for maintaining accuracy. Expertise requirements refer to the proper operation of DTs, that require specialized knowledge and skills in both the technology and application domains. Ensuring the security and privacy of data used and generated by DTs is paramount, given the sensitive nature of the data involved. In the remainder of the section, the mentioned challenges are discussed in more detail. 

\subsection{Technical and Setup Challenges}
The first challenge refers to the following issues:
\begin{itemize}
    \item interoperability and standardization,
    \item data acquisition and sensor setup,
    \item scalability,
    \item user-friendly interfaces.    
\end{itemize}
Ensuring interoperability and standardization is crucial for the effective implementation of DTs. Adopting standards like ISO/IEEE 11073 (X73) for personal health devices and MPEG-V for sensory devices facilitates seamless integration and data exchange, allowing various systems to work together without dealing with numerous proprietary data formats \cite{laamarti2019towards}. This interoperability is vital for achieving plug-and-play functionality, particularly in Digital Twin Coaching (DTC) systems, where standardized communication protocols enhance overall system effectiveness \cite{gamez2020digital}. Standardized protocols like IEEE X73 also improve the reliability and robustness of Human Digital Twins (HDTs), ensuring seamless interaction and data exchange \cite{lauer2024towards}.

Data acquisition and sensor setup are critical for DTs. Advanced sensors like 3D accelerometers and bio-sensors capture physiological data accurately, while IoT devices, including smart wearables and RFIDs, gather comprehensive health and environmental data. Proper placement and synchronization of these sensors ensure reliable data capture and real-time communication \cite{alsubai2023hybrid}.

Handling numerous Personalized Digital Twins (PDTs) requires significant resources, due to the complex interactions and frequent updates needed to reflect physical changes. Managing large data volumes and multiple participants is critical for success in broader applications \cite{sahal2022personal}.

Creating user-friendly interfaces is crucial for the effective use of DT systems. Simple and intuitive interfaces reduce the expert knowledge required significantly, making the technology more accessible to a wider audience. Incorporating features like clear visual feedback and compatibility with various devices enhances user interaction and engagement. Additionally, using avatars and virtual assistants can improve the user's understanding and trust in the system \cite{lauer2024towards}.

\subsection{Expertise requirements}
Expertise requirement challenges address primarily the issue of multidisciplinary collaboration. Successful implementation of Human Digital Twin Systems (HDTS) requires robust models addressing various human aspects, including physical, physiological, cognitive, and behavioral factors. This development necessitates a concerted, multidisciplinary effort, to synthesize these models into comprehensive HDTS applications. Such collaboration integrates expertise from different fields, to accelerate the deployment and effectiveness of~\cite{miller2022unified}.

\subsection{Data security and privacy}
Data security and privacy direct to the following issues:
\begin{itemize}
    \item data integrity and security,
    \item compliance with regulations,
    \item ethical data management.
\end{itemize}
Implementing DTs involves strict measures to ensure data security and privacy, focusing on data integrity and security, regulatory compliance, and ethical data management. Ensuring data integrity and security is critical. This involves using encryption, secure data storage, and protocols like SSL to protect data during transmission~\cite{alsubai2023hybrid}. Adhering to regulatory standards like GDPR ensures lawful data handling, and maintains user trust and compliance. Ethical data management requires transparency and user consent, ensuring clear information on the data use to build trust~\cite{lauer2024towards}.  

\subsection{Opportunities for future development}
The future development of digital twins presents numerous opportunities across various domains:
\begin{itemize}
    \item integration with advanced technologies,
    \item enhanced data analytics,
    \item advanced sensor technologies.
\end{itemize}
Advanced technologies such as AI and ML can enhance the capabilities of DTs in sports significantly. These technologies enable predictive analytics, real-time data processing, and more effective decision-making processes~\cite{gamez2020digital}. Another important domain is advanced data analytics, which can provide deeper insights into athletic performance, injury prevention, and recovery processes. By leveraging big data and sophisticated analytical tools, DTs can analyze vast amounts of data to identify patterns and trends, offering actionable insights to optimize training and performance~\cite{barricelli2020human}. The advancement of sensor technologies will enhance the accuracy and reliability of data collected by DTs. Innovations in sensors can lead to better monitoring of physical and environmental conditions, enabling more precise and comprehensive digital representations~\cite{gamez2021digital}. 

DTs offer a powerful tool to counteract the negative effects of a sedentary lifestyle. By creating personalized fitness programs, DTs can help individuals engage in regular physical activity tailored to their needs and conditions. These programs can be designed to fit into busy schedules, providing real-time feedback and adjustments to ensure the most effective exercise routines. For instance, a DT can monitor your daily activity levels and suggest exercises that can be done at your desk or during short breaks, promoting a more active lifestyle~\cite{gamez2020digital, laamarti2019towards}. In addition to enhancing general well-being, DTs play a crucial role in supporting amateur sports. Many amateur athletes lack access to professional coaching due to financial or geographical constraints. DTs can bridge this gap, by providing the guidance and expertise typically offered by real coaches. Systems like the AST, supported by DTs such as the AST-Monitor, can simulate the role of a coach, offering tailored training advice and feedback~\cite{fister2021trainer}.

\section{Conclusion}\label{sec:9}

This paper reviewed the role and impact of DTs in sports. Our review began by discussing the general concept of DTs. Through an SLR, we analyzed a collection of papers, to understand the current state of research and real-world applications of DTs. Four research questions were set in the beginning of the paper that are complete, as follows: 

\textbf{Answer to RQ1}: Our SLR revealed that individual sports, particularly \textbf{cycling}, \textbf{fitness}, and \textbf{football}, have seen the most significant support from DT technology. Other sports, like shooting, dancing, swimming, and gymnastics, have also seen the application of DTs, although to a lesser extent. Additionally, DTs have been applied in education, particularly physical education, providing valuable insights and enhancing the learning experience (see Section~\ref{sec:5}). 

\textbf{Answer to RQ2}: The maturity and practical implementation of DT technology in sports are varied. Some DT systems, such as \textbf{AST-Monitor} and \textbf{DTCoach}, are relatively mature and documented across multiple studies, indicating a higher level of development and application readiness. These systems demonstrate robust real-world applications with practical training and performance analysis benefits. However, many DT implementations are still in the prototype or experimental stages, with ongoing research needed to enhance their scalability, efficiency, and integration into everyday sports settings. This suggests that, while DT technology holds significant promise, widespread adoption and standardization are still in progress (see Section~\ref{sec:6}).

Our proposed taxonomy classifies DTs in sports according to their types, applications, and traits, providing a structured framework that can guide future research and development. This taxonomy helps to clarify the diverse roles that DTs can play in different sports disciplines and training scenarios (see Section~\ref{sec:5}).

\textbf{Answer to RQ3}: DTs enhance sports training significantly by providing detailed, real-time feedback, \textbf{personalized training programs}, and advanced \textbf{performance analysis}. Integrating data from sensors and wearable devices allows DTs to offer insights into an athlete's biomechanics, physiological responses, and overall performance. This allows for training interventions, rapid adjustments during training sessions, and more informed decision-making by coaches and athletes. For example, DTs can simulate various training scenarios, optimize techniques, and \textbf{prevent injuries} through detailed analysis. Using DTs in \textbf{strategic planning and competition simulation} further help refine tactics and improve overall performance outcomes (see Section~\ref{sec:7}).

\textbf{Answer to RQ4}: Implementing DT technology in sports presents several challenges. \textbf{Technical and setup challenges} include the need for standardized data acquisition methods, interoperability of different sensor technologies, and user-friendly interfaces. Ensuring the reliability and accuracy of data captured by various sensors is crucial for the effective functioning of DT systems. \textbf{Expertise requirements} pose another significant challenge, as the development and operation of DTs require multidisciplinary collaboration and specialized knowledge. \textbf{Data security and privacy} are also critical concerns, given the sensitive nature of the data involved in DT applications. Ensuring compliance with data protection regulations and maintaining ethical data management practices are essential for adopting DT technology in sports successfully (see Section~\ref{sec:8}).

Looking forward, the future development of DTs in sports presents numerous opportunities. Integrating advanced technologies such as artificial intelligence and machine learning can further enhance the capabilities of DTs. Improved data analytics and advancements in sensor technology will lead to more accurate and reliable digital models. These developments promise to refine the effectiveness of DTs, making them an integral part of sports training and performance optimization.
 
In summary, while the adoption of DTs in sports is still in its early stages, their potential to revolutionize the industry is immense. As technology evolves, DTs are poised to become a critical tool for athletes and coaches, driving advancements in training, performance monitoring, and strategic planning. It is important for DTs to be adopted in various sports, and to advance beyond the prototype stage into fully integrated systems. The ongoing research and innovation in this field will undoubtedly shape the future of sports, making it an exciting area for further exploration and development.

 \bibliographystyle{elsarticle-num} 
 \bibliography{cas-refs}

\end{document}

%% file: featuretable.tex
  \begingroup 
    \renewcommand*{\arraystretch}{1.5}%
    \definecolor{tabred}{RGB}{230,36,0}%
    \definecolor{tabgreen}{RGB}{0,116,21}%
    \definecolor{taborange}{RGB}{255,124,0}%
    \definecolor{tabbrown}{RGB}{171,70,0}%
    \definecolor{tabyellow}{RGB}{255,253,169}%
    \definecolor{tabviolet}{rgb}{0.6, 0.4, 0.8}
    \newcommand*{\redtriangle}{\textcolor{tabred}{\ding{115}}}%
    \newcommand*{\greenbullet}{\textcolor{tabgreen}{\ding{108}}}%
    \newcommand*{\greenone}{\textcolor{tabgreen}{\ding{182}}}%
    \newcommand*{\greencouple}{\textcolor{tabgreen}{\ding{183}}}%
    \newcommand*{\greenfew}{\textcolor{tabgreen}{\ding{184}}}%
    \newcommand*{\greenfewless}{\textcolor{tabgreen}{\ding{174}}}%
    \newcommand*{\greenseveral}{\textcolor{tabgreen}{\ding{185}}}%
    \newcommand*{\greenmany}{\textcolor{tabgreen}{\ding{186}}}%
    \newcommand*{\orangecirc}{\textcolor{taborange}{\ding{109}}}%
    \newcommand*{\violetsquare}{\textcolor{tabviolet}{\ding{110}}}%
    \newcommand*{\headformat}[1]{{\small#1}}%
    \newcommand*{\vcorr}{%
      \vadjust{\vspace{-\dp\csname @arstrutbox\endcsname}}%
      \global\let\vcorr\relax
    }%
    \newcommand*{\HeadAux}[1]{%
      \multicolumn{1}{@{}r@{}}{%
        \vcorr
        \sbox0{\headformat{\strut #1}}%
        \sbox2{\headformat{Complex Data Movement}}%
        \sbox4{\kern\tabcolsep\redtriangle\kern\tabcolsep}%
        \sbox6{\rotatebox{45}{\rule{0pt}{\dimexpr\ht0+\dp0\relax}}}%
        \sbox0{\raisebox{.5\dimexpr\dp0-\ht0\relax}[0pt][0pt]{\unhcopy0}}%
        \kern.75\wd4 %
        \rlap{%
          \raisebox{.25\wd4}{\rotatebox{45}{\unhcopy0}}%
        }%
        \kern.25\wd4 %
        \ifx\HeadLine Y%
          \dimen0=\dimexpr\wd2+.5\wd4\relax
          \rlap{\rotatebox{45}{\hbox{\vrule width\dimen0 height .4pt}}}%
        \fi
      }%
    }%
    \newcommand*{\head}[1]{\HeadAux{\global\let\HeadLine=Y#1}}%
    \newcommand*{\headNoLine}[1]{\HeadAux{\global\let\HeadLine=N#1}}%
    \noindent
    \begin{table}[htb]
    \begin{scriptsize}
    \begin{tabular}{%
      >{\bfseries}l|>{\quad}
      *{4}{c|}>{\quad}c|
      *{4}{c|}>{\quad}c|
      *{4}{c|}c%
    }%
      &
      \head{Nr. of papers} &
      \head{DTP} &
      \head{DTI} &
      \head{Predictive DTE} &
      &
      \head{Component} &
      \head{Product} &
      \head{System} &
      \head{Process} &
      &
      \head{Same look} &
      \head{Different details} &
      \head{Same behavior} &
      \head{Prediction \& information} &
      \\
      \hline
      \sbox0{S}%
      \rule{0pt}{\dimexpr\ht0 + 2ex\relax}%
      Aid Robots DT & \textcolor{tabgreen}{\bfseries 1} & \greenbullet & \greenbullet & \greenbullet && \redtriangle &  &  &  && & \violetsquare &  & \violetsquare \\
      Athlete Training Sys. & \textcolor{tabgreen}{\bfseries 1} & \greenbullet & \greenbullet & \greenbullet &&  &  & \redtriangle &  && & \violetsquare &  & \violetsquare \\
      AST Monitor & \textcolor{tabgreen}{\bfseries 3} & \greenbullet & \greenbullet & \greenbullet &&  & & \redtriangle &  &&  & \violetsquare &  & \violetsquare \\
      Conn. Footballer & \textcolor{tabgreen}{\bfseries 1} & \greenbullet & & && \redtriangle & & & && & \violetsquare & & \\
      Digital Athlete & \textcolor{tabgreen}{\bfseries 1} & \greenbullet & \greenbullet & \greenbullet &&  & \redtriangle &  &  &&  & \violetsquare &  & \violetsquare \\
      DT for Fitness & \textcolor{tabgreen}{\bfseries 1} & \greenbullet & \greenbullet & \greenbullet &&  & \redtriangle &  &  && & \violetsquare &  & \violetsquare \\
      DT for Gymnastics & \textcolor{tabgreen}{\bfseries 1} & \greenbullet & \greenbullet & \greenbullet && \redtriangle & &  &  && & \violetsquare &  & \violetsquare \\
      DT for swimmers & \textcolor{tabgreen}{\bfseries 1} & \greenbullet & \greenbullet & \greenbullet && \redtriangle & &  &  && & \violetsquare &  & \violetsquare \\
      DT in education & \textcolor{tabgreen}{\bfseries 1} & \greenbullet & \greenbullet & \greenbullet && \redtriangle & & &  &&  & \violetsquare &  & \violetsquare \\
      DT Model & \textcolor{tabgreen}{\bfseries 1} & \greenbullet & \greenbullet & \greenbullet && \redtriangle & &  &  &&  & \violetsquare &  & \violetsquare \\
      DTCoach & \textcolor{tabgreen}{\bfseries 3} & \greenbullet & \greenbullet & \greenbullet &&  &  & \redtriangle &  &&  & \violetsquare &  & \violetsquare \\ 
      DT Coaching & \textcolor{tabgreen}{\bfseries 1} & \greenbullet & \greenbullet & \greenbullet &&  & \redtriangle &  &  && & \violetsquare &  & \violetsquare \\
      MetaHuman & \textcolor{tabgreen}{\bfseries 1} & \greenbullet & \greenbullet & \greenbullet &&  & \redtriangle &  &  &&  & \violetsquare &  & \violetsquare \\
      Shooting DT & \textcolor{tabgreen}{\bfseries 1} & \greenbullet &  &  && \redtriangle &  & & &&  & \violetsquare & & \\
      SmartFit & \textcolor{tabgreen}{\bfseries 1} & \greenbullet & \greenbullet & \greenbullet &&  & \redtriangle &  &  && & \violetsquare  &   & \violetsquare \\
      Turtle DT & \textcolor{tabgreen}{\bfseries 1} & \greenbullet & &  && \redtriangle &  & & &&  & \violetsquare &  & \\
      \hline
      \multicolumn{1}{c|}{\bfseries Digital Twin} &
      \multicolumn{4}{c|}{\bfseries Type} &
      \multicolumn{5}{c|}{\bfseries Application} &
      \multicolumn{3}{c}{\bfseries Trait}
      \\
    \end{tabular}%
    \caption{Taxonomy of the digital twin in sport.}
    \label{fig:taxonomy}
    \end{scriptsize}
    \end{table}
    \kern19.5mm 
  \endgroup